\documentclass[twocolumn,aps,prb,amsfonts]{revtex4}
\usepackage{graphicx}
\usepackage{psfrag, color}
\usepackage{verbatim}
\usepackage{amsmath}

\begin{document}

\title{Exploration of the Limits to Mobility in Two-Dimensional Hole Systems in GaAs/AlGaAs Quantum Wells}

\author{J. D. Watson$^{1,2}$, S. Mondal$^{1,2}$, G. Gardner$^{2,3}$, G. A. Cs\'{a}thy$^1$, M. J. Manfra$^{1,2,3,4}$\footnote[1]{mmanfra@purdue.edu}}

\affiliation{${}^1$ Department of Physics\\
${}^2$ Birck Nanotechnology Center\\
${}^3$School of Materials Engineering\\
${}^4$ School of Electrical and Computer Engineering\\
Purdue University,
West Lafayette, IN 47907\\
}

\begin{abstract}
We report on the growth and electrical characterization of a series of two-dimensional hole systems (2DHSs) used to study the density dependence of low temperature mobility in 20 nm GaAs/AlGaAs quantum wells.  The hole density was controlled by changing the Al mole fraction and the setback of the delta-doping layer.  We varied the density over a range from 1.8 $\times$ 10$^{10}$ cm$^{-2}$ to 1.9 $\times$ 10$^{11}$ cm$^{-2}$ finding a nonmonotonic dependence of mobility on density at T = 0.3 K.  Surprisingly, a peak mobility of 2.3 $\times$ 10$^6$ cm$^2$/Vs was measured at a density of 6.5 $\times$ 10$^{10}$ cm$^{-2}$ with further increase in density resulting in reduced mobility.  We discuss possible mechanisms leading to the observed non-monotonic density dependence of the mobility.  Relying solely on interface roughness scattering to explain the observed drop in mobility at high density requires roughness parameters which are not consistent with measurements of similar electron structures.  This leaves open the possibility of contributions from other scattering mechanisms at high density.  
\end{abstract}

\maketitle
Two-dimensional hole systems (2DHSs) on (001) oriented GaAs offer an interesting alternative to the more widely studied two-dimensional electron systems (2DESs).  2DHSs on (001) GaAs have effective masses roughly 4.5 to 7.5 times larger\cite{lu2008cyclotron,zhu2007density,khannanov2007dependence} than that in corresponding 2DESs which increases the importance of Coulomb interactions relative to the kinetic energy resulting in enhancement of importance of many-body effects.  In addition, the p-wave symmetry of the valence band in GaAs leads to a much reduced hyperfine coupling of hole spins to the atomic nuclei which makes them an exciting alternative to electrons for quantum dot spin-based qubits\cite{brunner2009coherent,gerardot2008optical,bulaev2005spin}.  The presence of spin-orbit coupling and light/heavy hole mixing in the valence band of GaAs also allows extensive band structure engineering\cite{eisenstein1984effect,winklerbook,habib2009spin}. This feature has been exploited to alter the nature of groundstates in the quantum Hall regime.\cite{manfra2007impact,koduvayur2011effect} 

Here we describe our efforts to understand the limits to low temperature mobility for (001) 2DHSs.  Continued improvement in 2DHS quality is motivated by the well-established paradigm for 2DESs that increased low-temperature mobility often leads to the observation of new correlated groundstates\cite{pfeiffer2003therole}.  Historically, improvement to the low temperature mobility of 2DHSs has lagged behind that of 2DESs due to the lack of a p-type dopant in GaAs that does not diffuse or segregate significantly at typical molecular beam epitaxy (MBE) growth temperatures $\sim$ 635 $^{\circ }$C.  Si can act as a low-diffusivity acceptor on the (311)A face of GaAs, but subsequent transport experiments are known to be complicated by a significant mobility anisotropy due to surface corrugation\cite{Her_San_JAP_1994}.  However, recent use of low diffusivity carbon doping (C-doping)\cite{Reu_Wie_RSI_1999,Man_Pfe_APL_2005,Ger_Sch_APL_2005} has rapidly led to low temperature mobilities $>$ 10$^6$ cm$^2$/Vs without the accompanying transport anisotropy.  Purely from a growth standpoint, then, there does not appear to be any reason why low temperature hole mobilities should not approach that of electrons once scaled by the appropriate effective mass. Presently it is widely believed that uniformly distributed ionized background impurities limit the mobility in the best 2DESs \cite{Hwa_Das_PRB_2007}. However, the highest hole mobility reported to date\cite{watson2011scattering} of 2.6 $\times$ 10$^{6}$ cm$^2$/Vs is still about a factor of two lower than record mobility 2DESs grown in the same MBE chamber\cite{pfeiffer2003therole} once the heavy hole to electron effective mass ratio of 0.4m$_e$ : 0.067m$_e$ is taken into account.  The question then remains, if sufficiently reducing background impurities\cite{Hwa_Das_PRB_2007} is the main obstacle to reaching an electron mobility of 100 $\times$ 10$^6$ cm$^2$/Vs, what are the key ingredients to a hole mobility of 15 $\times$ 10$^6$ cm$^2$/Vs?

\begin{figure}[t]
\includegraphics[width=3in]{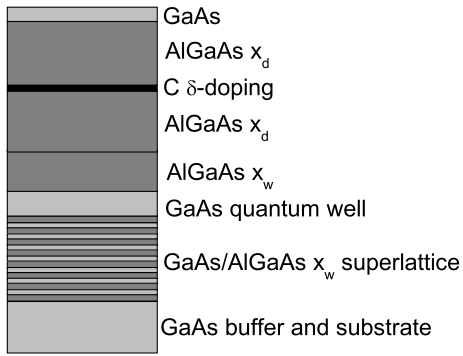}
\caption{Layer structure of devices in this experiment.  Note the use of two different Al mole fractions $x_w$ and $x_d$ in some of the devices as indicated in Table \ref{summary}.}
\label{structure}
\end{figure}

\begin{figure}[t]
\includegraphics[width=3in]{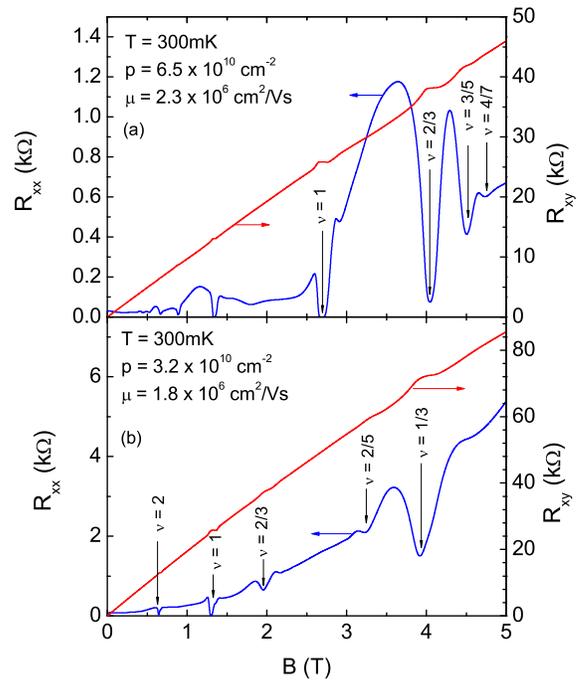}
\caption{(Color online) Magnetotransport at \textit{T} = 300 mK after illumination with a red LED of (a) peak mobility sample and (b) low density sample that exhibits many nascent fractional QHE features.}
\label{comb_Hall}
\end{figure}

In order to answer this question, we have begun to explore the impact of varying structural parameters on the resulting mobility.  Samples in this work were grown in a customized Veeco GenII MBE which has recently achieved {\it electron} mobilities $>$ 20 $\times$ 10$^6$ cm$^2$/Vs and extremely large excitation gaps for the fragile $\nu$=5/2 fractional quantum Hall state.  C-doping was performed with a carbon filament capable of producing a doping rate of 2.8 $\times$ 10$^{10}$ cm$^{-2}$/sec at a total power (including parasitic dissipation) of $\sim$ 150 W.\cite{pfeiffercarbon}  
In this experiment, we utilized a 20 nm quantum well situated 190 nm below the surface and asymmetrically $\delta$-doped from above at a setback $d$ of 80, 110, or 140 nm.  The Al mole fraction $x$ was also varied between 0.07 and 0.45 to allow further tuning of the 2DHS density.  Table \ref{summary} summarizes the structures grown in the experiment, and Fig. \ref{structure} shows the epilayer design.  
\begin{table}
\caption{Summary of structural parameters including $\delta$-doping setback distance $d$, Al mole fraction around the dopants $x_{d}$, Al mole fraction surrounding the quantum well $x_{w}$, 2DHS density $p$, and \textit{T} = 300 mK mobility after illumination $\mu$.}
\begin{tabular}{  l  c  c  c  c  c }
\hline
\hline
Sample          & $d$ & $x_{d}$ & $x_{w}$ & $p$                 & $\mu$  \\ 
                & nm  &              &                 & 10$^{11}$cm$^{-2}$  & 10$^6$ cm$^2$/Vs \\ \hline
\textbf{1}      & 80  & 0.24         & 0.24            & 1.1                 & 1.2 \\
\textbf{2}      & 80  & 0.24         & 0.24            & 0.98                & 1.4 \\
\textbf{3}      & 80  & 0.45         & 0.45            & 1.9                 & 0.55 \\
\textbf{4}      & 80  & 0.10         & 0.10            & 0.32                & 1.8 \\
\textbf{5}      & 80  & 0.35         & 0.35            & 1.4                 & .80 \\
\textbf{6}      & 80  & 0.20         & 0.20            & 0.80                & 1.6 \\
\textbf{7}      & 80  & 0.07         & 0.07            & 0.18                & 1.3 \\
\textbf{8}      & 110 & 0.10         & 0.10            & 0.29                & 1.5 \\
\textbf{9}      & 140 & 0.10         & 0.10            & 0.23                & 1.4 \\
\textbf{10}      & 110 & 0.24         & 0.24            & 0.70                & 1.6 \\
\textbf{11}     & 80  & 0.16         & 0.16            & 0.65                & 2.3 \\
\textbf{12}     & 110 & 0.13         & 0.13            & 0.36                & 1.8 \\
\textbf{A}      & 80  & 0.45         & 0.16            & 1.7                 & 0.73 \\
\textbf{B}      & 80  & 0.45         & 0.24            & 1.5                 & 0.78 \\
\textbf{C}      & 80  & 0.35         & 0.16            & 1.34                & 1.3 \\
\textbf{D}      & 80  & 0.35         & 0.24            & 1.30                & 1.1 \\
\hline
\hline
\end{tabular}
\label{summary}
\end{table}
Square samples were prepared using InZn contacts annealed at 430 $^{\circ}$C for 15 minutes in H$_2$/N$_2$ forming gas.  Characterization was performed in the dark and after illumination with a red LED at \textit{T} = 300 mK using standard lock-in techniques, and the density was determined from quantum Hall effect (QHE) minima.  Illumination typically resulted in $\sim$ 3-5$\%$ increase in density and as much as a 27$\%$ increase in mobility for low density samples.  Transport data also showed a qualitative improvement after illumination, indicating that illumination increases the homogeneity of the 2DHS and has a favorable impact on the screened disorder potential.  Figure \ref{comb_Hall} shows transport data of the highest mobility sample and a low density sample; the number of nascent fractional QHE features attest to the sample quality.     
\begin{figure}[t]
\includegraphics[width=3in]{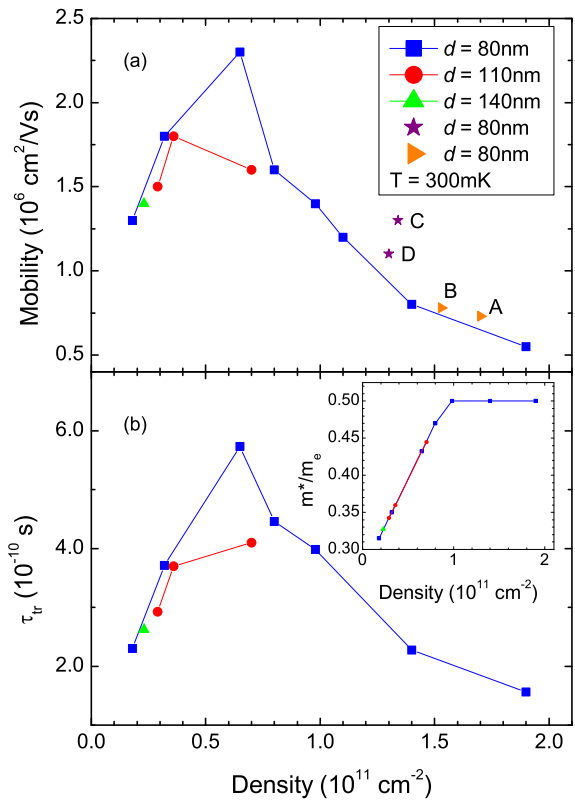}
\caption{(Color online) (a) \textit{T} = 300 mK mobility after illumination with a red LED as a function of density for various dopant setback distances $d$.  Solid lines are guides to the eye.  For fixed $d$ the density was controlled by varying the Al mole fraction $x$.  Samples were grown in random order to avoid continued machine clean-up from skewing the observed trend in mobility.  Samples A-D were grown with varying $x$ at fixed $p$ to test the effect of alloy and interface roughness scattering on the mobility (see text).  (b) Transport lifetime estimated as a function of density. Inset: Effective mass for our structures as a function of density extrapolated from refs. [\onlinecite{lu2008cyclotron,zhu2007density}].}
\label{main_result}
\end{figure}

Fig. \ref{main_result}a shows the measured mobility as a function of density for various values of $d$.  We note that remote ionized impurity (RI) scattering does not play a significant factor in limiting the mobility since within experimental uncertainty there is no meaningful difference between the mobility at different values of $d$ for the same density.  However, increased $d$ should allow these samples to be gated to ultra-low densities before RI scattering begins to cause the mobility to rapidly drop off with further decreased density\cite{watson2011scattering}.  
%
%
%
The most interesting feature of the data in Fig. \ref{main_result}a, however, is the strongly non-monotonic dependence of the mobility on density.  For 2DESs in this density range with such a large value of $d$, one would expect the mobility to monotonically increase with density\cite{pfeiffer2003therole,Hwa_Das_PRB_2007,umansky1997extremely,pfeiffer1989electron} following a power law dependence $\mu \propto p^{\alpha}$ where $\alpha$ $\sim$ 0.6 - 0.8 with ionized background impurity (BI) scattering being the dominant scattering mechanism.  In analyzing our results, we first note that the effective mass is known to vary throughout the density range of our samples due to the valence band non-parabolicity arising from light- and heavy-hole band mixing.  By performing a linear fit to cyclotron resonance data on 2DHSs in (001) 20 nm quantum wells in refs. [\onlinecite{lu2008cyclotron,zhu2007density}] and assuming the cyclotron mass plateaus at 0.5m$_e$ at high density we estimate the transport lifetime for our structures as shown in Fig. \ref{main_result}b.
%
%
The transport lifetime, however, follows the same non-monotonic behavior as the mobility which indicates a competition between different scattering mechanisms throughout the density range of our experiment in addition to the changing mass.  

To shed further light on possible scattering mechanisms, we have performed a series of scattering calculations including the effects of BI, RI, alloy, and interface roughness (IR) scattering.  We follow the derivation of the transport relaxation time in [\onlinecite{bastardbook}] which assumes \textit{T} = 0 and neglects intersubband scattering, multiple scattering events, and correlation between ionized impurities.  This simple calculation is intended to elucidate the expected trend of the mobility as the density is increased and determine if scattering mechanisms dominant in 2DESs can qualitatively explain our observations.  More sophisticated calculations have been made by S. Das Sarma and coworkers.\cite{dassarma1999charged,dassarma2000calculated,Hwa_Das_PRB_2007}  Transport relaxation times are calculated individually and then the total mobility is calculated using Mathiessen's rule.  For BI and RI scattering the transport lifetime is given by 
\begin{multline}
\frac{1}{\tau_{tr}(\epsilon _F)} = \frac{m^*}{\pi \hbar^3}\sum _i \int _0^{\pi} \mathrm{d}\theta(1-\cos(\theta)) \\
\times \left[\frac{2\pi e Z_ie_i}{4\pi\epsilon\left(q + q_{TF}g_s(\textbf{q})\right)}\right]^2 \int_{-\infty}^{\infty}\mathrm{d}z N_i(z)g_{imp}^2(q,z)
\label{tauBIRI}
\end{multline}
where $m^*$ is the hole effective mass (as estimated in Fig. \ref{main_result}b), $\hbar$ is the reduced Planck constant, $\theta$ is the scattering angle, $Z_ie_i$ is the impurity charge, $\epsilon$ is the dielectric constant of the semiconductor, $\textbf{q}$ is the scattering vector, $k_F$ is the Fermi wavevector, $q_{TF}$ is the Thomas-Fermi screening wavevector, $N_i(z)$ is the i$^{\mathrm{th}}$ impurity distribution, and the form factors are given by
\begin{equation}
g_s(\textbf{q}) = \int \chi ^2(z) \chi ^2(z') \mathrm{exp}(-q|z - z'|) \mathrm{d}z\mathrm{d}z'
\label{gs}
\end{equation}
\begin{equation}
g_{imp}(q,z) = \int \chi ^2(z')\mathrm{exp}(-q|z'-z|)\mathrm{d}z'
\label{gimp}
\end{equation}
where $\chi(z)$ is the self-consistently calculated\cite{nextnano} envelope function in the effective mass approximation.  For the BI calculation we use a three-dimensional impurity concentration $N_{3D}$ as a fitting parameter and find the best agreement with the experimental data for $N_{3D}$ = 2 $\times$ 10$^{13}$ cm$^{-3}$.  We use a remote impurity sheet concentration $N_{RI}$ equal to the hole concentration $p$.
A more realistic value of $N_{RI}$ could also include some of the ionized impurities due to the surface compensation; however, we assume a simple parallel-plate capacitor model of the surface-$\delta$-layer charge and thus neglect the surface compensation contribution to $N_{RI}$.
This neglect of charge due to surface compensation is typical in these types of calculations.\cite{Hwa_Das_PRB_2007,lee1983low,stern1983doping}  For our purposes, though, the exact value of $N_{RI}$ is not important since it will not change the qualitative dependence of the RI-limited mobility as $p$ is varied.  

To calculate alloy scattering we use the virtual crystal approximation with a square well potential limited over a spherical range\cite{harrison1976alloy} which is independent of temperature in 2D systems\cite{chattopadhyay1985alloy}.  The alloy limited relaxation lifetime is unscreened due to its short range nature and given by\cite{bastardbook,li2003direct}
\begin{equation}
\frac{1}{\tau_{alloy}(\epsilon_F)}=\frac{4\Omega ^2m^*U^2x(1-x)}{a^3\hbar ^3}\int\limits_{barrier}\chi ^4(z)\mathrm{d}z
\label{taualloy}
\end{equation}
where $a$ = 0.565 nm is the lattice constant of the compound semiconductor, $\Omega$ is the volume of the scattering potential given by $\Omega = (4/3)\pi r^3$, and $r=(\sqrt{3}/4)a$ is the nearest-neighbor separation.  There is a broad range of estimates of the magnitude of the scattering potential $U$ in the literature\cite{li2003direct}, ranging from 0.12 to 1.56 eV.  We take $U$= 1 eV (as suggested in ref. [\onlinecite{bastardbook}]) as a rough estimate.  

To examine the possible effect of interface roughness scattering, we employ a simple model which makes use of the Fang-Howard variational wavefunction and associated potential\cite{fang1966negative} which takes the distortion of the wavefunction with increased density into account.  In this model, the IR scattering rate is given by\cite{ando1982electronic,ando1982selfconsistent}
\begin{multline}
\frac{1}{\tau_{IR}(\epsilon_F)} = \left(\frac{\Delta \Lambda e^2 p}{2\epsilon}\right)^2\frac{m^*}{\hbar^3}\int_0^{\pi}\left(\frac{q}{q + g_s(\textbf{q})q_{TF}}\right) ^2 \\
\times (1-\cos \theta)\mathrm{exp}(-\Lambda^2 q^2/4)\mathrm{d}\theta
\label{IR_FH}
\end{multline}
where the wavefunction used to calculate $g_s$ is
\begin{displaymath}
\chi(z) = \left\{
   \begin{array}{lr}
     \frac{1}{\sqrt{2}}b^{3/2}ze^{-bz/2} & z > 0\\
     0 & z \leq 0
   \end{array}
  \right.
\end{displaymath}
where the variational parameter is\cite{daviesbook,ando1982electronic}
\begin{equation}
b = \left(\frac{33m^*e^2p}{8\hbar^2\epsilon}\right)^{1/3}
\label{variational}
\end{equation}
We take one monolayer roughness height to be a reasonable estimate and thus set $\Delta$ = 0.2825 nm and use $\Lambda$ as a fitting parameter with the result that $\Lambda$ = 6 nm.  

\begin{figure}[t]
\centering
\includegraphics[width=3in]{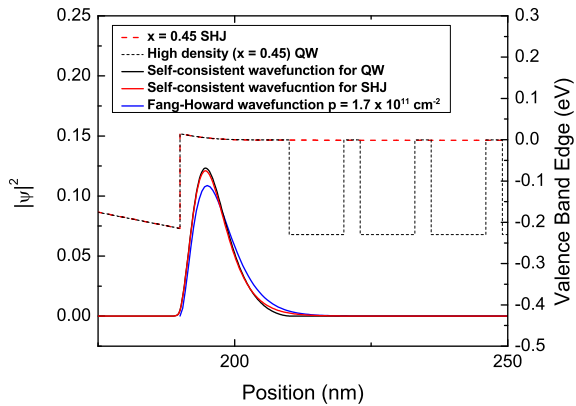}
\caption{(Color online) Dashed lines show a comparison of the self-consistently calculated valence band edges (dashed lines) for the high density sample \# 3 and and a single heterojunction sample with $x$ = 0.45.  Solid lines show a comparison of the self-consistently calculated wavefunction for sample \# 3 and the Fang-Howard variational wavefunction.}
\label{FH_Justification}
\end{figure}

As a justification for using the Fang-Howard wavefunction to model our asymmetric quantum well system, we show in Fig. \ref{FH_Justification} a comparison of the self-consistently calculated valence band edges for quantum well (QW) sample \# 3 and a single heterojunction (SHJ) along with the self-consistently calculated wavefunction for the QW structure and the Fang-Howard wavefunction which is often taken as an approximation of the wavefunction in SHJ structures.
The band edges show that the bottom barrier of the QW changes the confining potential very little, and the high density samples (where IR scattering could be important) can therefore be approximated by the Fang-Howard model. 

\begin{figure}[t]
\centering
\includegraphics[width=3in]{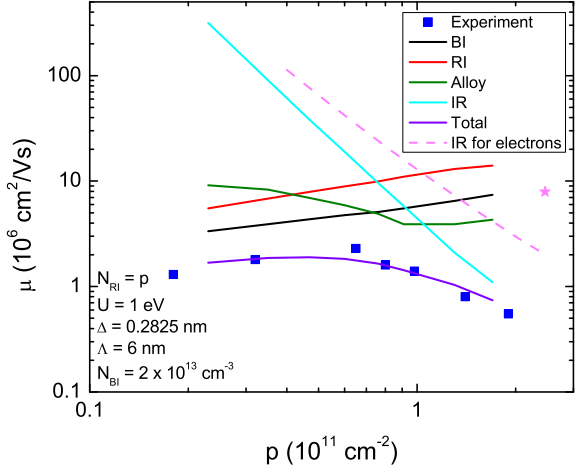}
\caption{(Color online) Comparison of $d$ = 80 nm experimental data with mobility calculations.  $N_{BI}$ and $\Lambda$ are used as free parameters to obtain a good fit to data.  Pink star represents SHJ 2DES grown during this experiment.}
\label{theory_result_FH_forced}
\end{figure}

The results of our calculations are compared with the $d$ = 80 nm experimental results in Fig. \ref{theory_result_FH_forced}. It is clear that even with a changing effective mass and wave function profile the BI and RI limited mobilities steadily increase with increasing density and therefore cannot account for the drop in mobility at high density.  The exact contributions of alloy and interface roughness scattering are initially less clear, however.  We will address alloy disorder first.

If $U$ is large enough, alloy scattering could conceivably contribute to the drop in mobility seen in the experimental data.  Before continuing, it should be noted that the slight increase in the calculated alloy-limited mobility at high density is simply due to the saturation of the effective mass as shown in the inset of Fig. \ref{main_result}b. We have repeated the calculations (not shown) without forcing the mass to plateau at 0.5m$_e$, but even with a mass as high as 0.7m$_e$ at high density the alloy limited mobility does not appear to be limiting the total mobility.  To test the contribution of alloy scattering we grew a series of four test structures (labeled A-D in Fig. \ref{main_result}a) in which $x_{d}$, the Al concentration starting 25 nm above the quantum well (e.g. around the $\delta$-doping layer), was kept fixed to keep the density constant while $x_{w}$, the Al concentration around the quantum well, was varied between $x_w$ = 0.16 and $x_w$ = 0.24. 
%
Samples A, B, and 3 ($x_d$ = 0.45) suggest that $x_w$ has no impact on the mobility, though there is scatter in the resulting density which we attribute to wafer-to-wafer variation and possible variation in the illumination.  Samples C, D, and 5 ($x_d$ = 0.35), however, suggest that increased $x_w$ does cause the mobility to decrease somewhat.  Most importantly, this is the opposite trend that would be expected if alloy scattering per se were limiting the mobility.  Our calculations for the test structures (not shown) and ref. [\onlinecite{ando1982selfconsistent}] predict that the alloy-limited mobility would increase for increased $x_w$ since as $x_{w}$ is increased for fixed density the wavefunction is more confined.  This in turn causes the integral of $\chi^4$ to decrease faster than the $x(1-x)$ term increases in equation (\ref{taualloy}) resulting in a decrease in the alloy scattering rate for increased $x_w$.  The results from this set of structures is consistent, however, with the theory that Al getters impurities\cite{pfeiffer2003therole}, thus an increase in $x_w$ would locally increase $N_{BI}$ and the associated scattering.  Regardless, samples C, D, and 5 show that the negative side effects of increasing $x_w$ are not enough to explain the data of Fig. \ref{main_result}.  If the increase in $x_w$ was dominating the mobility we would expect test structures A and C to have significantly higher mobilities than the peak mobility sample \#11 due to the higher hole density of the test structures.  This, however, is clearly not the case. 

\begin{figure}[t]
\centering
\includegraphics[width=3in]{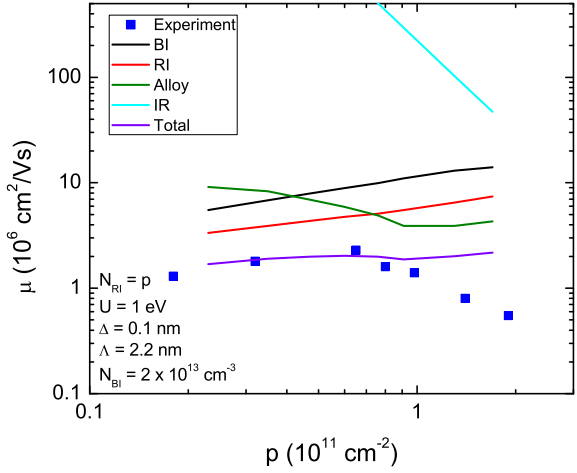}
\caption{(Color online) Comparison of $d$ = 80 nm experimental data with mobility calculations.  $\Delta$ and $\Lambda$ are varied in order for mobility calculations to obtain agreement with an electron structure grown during this experiment (see text).}
\label{FH_consistent}
\end{figure}

Finally, our fit seems initially to indicate that IR scattering is limiting the mobility at high density.  However, whenever parameters can be freely adjusted caution must of course be exercised to obtain physically meaningful results.  The dashed pink line in Fig. \ref{theory_result_FH_forced} shows the IR limited mobility for a SHJ 2DES in the Fang-Howard model while the pink star shows a 2DES SHJ structure with $x$ = 0.35 grown during the course of this experiment.  Evidently the Fang-Howard calculation overestimates the IR scattering by at least a factor of four.  Repeating our self-consistent calculation for BI, RI, and alloy scattering in this 2DES SHJ using the impurity concentrations and alloy potential listed in the inset of Fig. \ref{theory_result_FH_forced} we find that the IR-limited mobility at a density of 2.4 $\times$ 10$^{11}$ cm$^{-2}$ would have to be 86 $\times$ 10$^6$ cm$^2$/Vs to fit the measured total mobility of 7.9 $\times$ 10$^6$ cm$^2$/Vs.  To get such a high IR-limited mobility we are forced to set $\Delta$ = 0.1 nm and $\Lambda$ = 2.2 nm.  Figure \ref{FH_consistent} shows the result of our calculation for the hole structures using these smaller roughness parameters.  
  
With these reduced roughness parameters there is no longer a good fit to the hole data at high density as the IR term makes almost no contribution to the total mobility, though we still obtain a good fit at low to medium density.  We conclude that our crude model of interface roughness scattering cannot simultaneously account for our experimental data in both electrons and holes and are thus hesitant to conclude that interface roughness scattering is the dominant source of our drop in mobility at high density.  Similar discrepancies between electron and hole data have been noted in ref. [\onlinecite{chen2012fabrication}].  

Another possible scattering mechanism that must be kept in mind at high density is scattering between the electric subbands of the quantum well which is known to degrade the mobility in high density 2DESs.\cite{stormer1982observation}  To estimate the possibility of such scattering, we use a finite square well with a barrier height of 230 meV and an effective mass of 0.5m$_e$, which corresponds to our highest density sample. This estimate results in an energy spacing of 5.0 meV between the heavy hole ground and first excited state.  Assuming a light hole mass along the (001) direction\cite{daviesbook,winklerbook} of 0.090m$_e$ the spacing between the heavy hole and light hole ground states is 6.4 meV.  In both cases this energy spacing is significantly larger than the Fermi energy $E_F = \frac{\pi \hbar^2 p}{m*} \sim 0.9$ meV which precludes a significant contribution from intersubband scattering between the electric subbands.

\begin{figure}[t]
\includegraphics[width=3in]{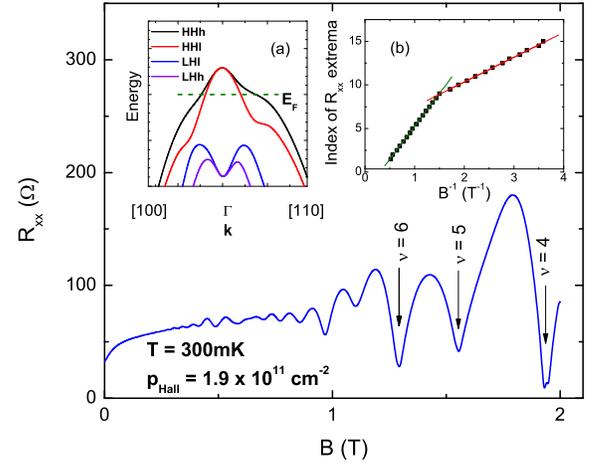}
\caption{(Color online) Shubnikov-de Haas oscillations of high density $d$ = 80 nm, $x$ = 0.45 device.  Inset (a): Sketch of the spin-split heavy hole and light hole ground states in a quantum well.  Inset (b): Index of extrema in R$_{xx}$ vs. B$^{-1}$.  The high field slope gives the total density of 1.8 $\times$ 10$^{11}$ cm$^{-2}$, and the low field slope gives the lighter sub-band density of 7 $\times$ 10$^{10}$ cm$^{-2}$ while the difference in the two gives the second sub-band density of 1.1 $\times$ 10$^{11}$ cm$^{-2}$. }
\label{SdH}
\end{figure}
Next, we note the presence of beating in the Shubnikov-de Haas oscillations in Fig. \ref{SdH} which is indicative of B = 0 spin-splitting.  Such spin-splitting is known to occur in structurally-asymmetric devices\cite{eisenstein1984effect,habib2009spin,stormer1983energy} due to Rashba spin-orbit coupling\cite{winklerbook}.  We sketch the qualitative effect of this splitting in inset (a) of Fig. \ref{SdH}. As the 2DHS density is increased, the electric field (and hence spin splitting) in the well is also increased.  Furthermore, it is know that the presence of a parallel channel can result in a Hall density different from the sum of the subband densities and a measured mobility different from that of either subband even in the absence of intersubband scattering.  In our case, we assume that the two parallel channels are non-interacting B = 0 spin-split subbands of the heavy hole ground state.      
%
%
%
%
%
%
   The measured Hall density $p_{Hall}$ and mobility $\mu_{Hall}$  in the absence of intersubband scattering are given by\cite{daviesbook}
\begin{equation}
p_{Hall} = \frac{(p_1\mu_1 + p_2\mu_2)^2}{p_1\mu_1^2 + p_2\mu_2^2}
\label{p_eff}
\end{equation}
\begin{equation}
\mu_{Hall} = \frac{p_1\mu_1^2 + p_2\mu_2^2}{p_1\mu_1 + p_2\mu_2}
\label{mu_eff}
\end{equation}
\begin{figure}[t]
\includegraphics[width=3in]{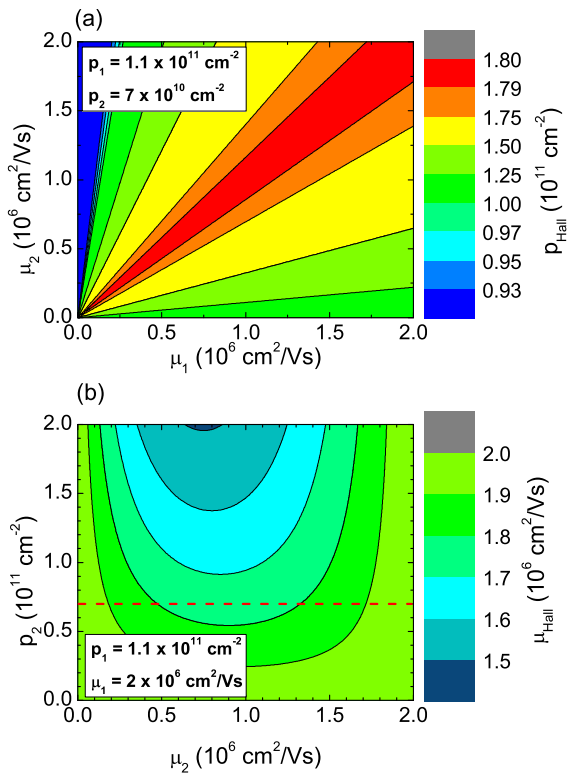}
\caption{(Color online) (a) Estimate of the Hall density expected from Eq. \ref{p_eff} using the measured subband densities $p_1 = 1.1 \times 10^{11}$ cm$^{-2}$ and $p_2 = 7 \times 10^{10}$ cm$^{-2}$.  (b) Estimate of the expected measured mobility $\mu_{Hall}$ if the high density subband $p_1 = 1.1 \times 10^{11}$ cm$^{-2}$ has a high mobility $\mu_1 = 2 \times 10^6$ cm$^2$/Vs.  For the second subband density (dashed red line) $p_2 = 7 \times 10^{10}$ cm$^{-2}$ measured in Fig. \ref{SdH} the expected measured mobility $\mu_{Hall} \geq 1.75 \times 10^6$ cm$^2$/Vs.}
\label{parallel}
\end{figure}
where $\mu_{1(2)}$ and $p_{1(2)}$ are the mobility and density, respectively, of the first (second) subband.  Figure \ref{parallel}a illustrates the Hall density as a function of the subband mobilities  in our peak density sample (sample \# 3) predicted by Eq. \ref{p_eff} using the subband densities extracted in Fig. \ref{SdH}.  It is clear from Fig. \ref{parallel}a that in order to measure a Hall density $\sim$ 1.8 $\times$ 10$^{11}$ cm$^{-2}$ the subband mobilities should be comparable, though the high density subband should have a slightly higher mobility.  In order to estimate the effect of the presence of two subbands on the measured mobility, we therefore assume that the high density subband is also the high mobility subband. In order to determine if the presence of the lower mobility subband could by itself account for the drop in mobility seen in Fig. \ref{main_result}, we assume that the high mobility subband is unchanged from the peak total mobility value ($\sim 2 \times 10^6$ cm$^2$/Vs) at low density.  Figure \ref{parallel}b shows what we would thus expect to measure as a function of density and mobility in the low mobility subband if the high mobility subband has a density $p_1 = 1.1 \times 10^{11}$ cm$^{-2}$ as we measure in Fig. \ref{SdH}.  For the measured second subband density of $p_2 = 7 \times 10^{10}$ cm$^{-2}$ (dashed red line) we see that this parallel subband effect would not decrease the measured mobility below $\sim$ 1.75 $\times$ 10$^6$ cm$^2$/Vs.  We therefore conclude that the presence of a second, possibly low mobility B = 0 spin-split subband cannot explain our observed drop in mobility at high density in the absence of intersubband scattering.

A final possible mechanism for the observed drop in mobility at high density is intersubband scattering between the spin-split subbands of the heavy hole ground state of the quantum well.  The question remains, however, whether or not there exists a potential capable of coupling the spin-split sub-bands and causing back-scattering.  Such scattering is typically neglected in theoretical calculations of the mobility due to the assumed lack of a significant spin-flip mechanism\cite{walukiewicz1985hole}, though intersubband hole-hole scattering in inversion-asymmetric structures is not without precedent.\cite{hwang2003temperature} At this time more theoretical work is needed to resolve the relative contributions of the different scattering mechanisms.

In conclusion, we have performed an experimental study of the density dependence of mobility in C-doped (001) GaAs/AlGaAs quantum wells by varying the dopant setback $d$ and Al mole fraction $x$. The mobility was seen to depend non-monotonically on the density.  At low density the mobility increased with density.  The \textit{T} = 300 mK mobility was found to peak at a value of 2.3 $\times$ 10$^6$ cm$^2$/Vs at a density of 6.5 $\times$ 10$^{10}$ cm$^{-2}$.  This 2DHS mobility is among the highest ever reported.  Increasing the density further, however, resulted in a sharp drop in mobility.  Scattering calculations indicate that background ionized impurities and remote ionized impurities will not lead to a decrease in mobility at high density even with a changing effective mass, and alloy scattering cannot account for all of our experimental results from various test structures.  Interface roughness scattering contributions remain unclear due to the difficulty in obtaining physically reasonable roughness parameters that predict both electron and hole mobilities.  Beating in the Shubnikov-de Haas oscillations in our high density samples is indicative of zero-field spin-splitting which leaves open the possibility of an intersubband scattering contribution to the mobility.  Further theoretical work is needed to determine the mechanism and magnitude of such a contribution.

\section*{Acknowledgment}
\vspace*{-10pt}
JDW would like to thank S. Birner, C. R\"{o}ssler, and T. Feil for helpful discussions regarding calculations with Nextnano3.  JDW is supported by a Sandia Laboratories/Purdue University Excellence in Science and Engineering Fellowship. MJM acknowledges support from the Miller Family Foundation.  The MBE growth and transport measurements at Purdue are supported by the U.S. Department of Energy, Office of Basic Energy Sciences, Division of Materials Sciences and Engineering under Award  DE-SC0006671.
%
\bibliography{Complete_Database}

\begin{thebibliography}{39}
\bibitem[{pfe()}]{pfeiffercarbon}
\bibinfo{howpublished}{MJM thanks Ken West and Loren Pfeiffer from Princeton
  University for sharing their carbon source design upon which ours is based.}
\end{thebibliography}


\end{document}